\documentclass[preprint,onecolumn,showpacs,preprintnumbers,amsmath,amssymb]{revtex4}
\usepackage{bm} \usepackage{graphics} \usepackage[pdftex]{graphicx} 
\DeclareMathOperator{\sign}{sign} \usepackage{ulem}

\begin{document} \frenchspacing

\title{The symmetry and magnetoelectric effects \\ in garnet crystals and films}

\author{A. I. Popov$^{1,2,3}$, D. I. Plokhov$^1$, and A. K. Zvezdin$^{1,3}$ \medskip}

\affiliation{$^1$A. M. Prokhorov General Physics Institute of Russian Academy of Sciences, \\ 38 Vavilov Str., 119991, Moscow, Russia \medskip \\ $^2$National Research University of Electronic Technology \\ 5 Pas. 4806, 124498, Zelenograd, Moscow, Russia \medskip \\ $^3$Moscow Institute of Physics and Technology \\ 9 Institutsky Per., 141700, Dolgoprudny, Moscow region, Russia \medskip}

\date{\today} 

\begin{abstract}
    The magnetoelectricity of garnets is considered by means of a symmetry and quantum mechanical combined analysis. It is shown, that the magnetoelectric effect is not realized in most garnets although the necessary condition of the crystal magnetic structure antisymmetry in them is held at low temperatures. Nevertheless, the effect can be observed in some garnets as well as other odd effects, namely, piezomagnetic effect, magnetic field evoked piezoelectric one, etc. It is also discovered that magnetic fields can induce specific antiferroelectric structures in garnet crystals and produce electric polarization in epitaxial films. The polarization can also be caused in a bulk crystal by an inhomogeneous magnetic field.
\end{abstract}


\maketitle


\section{Introduction} \label{sec1}

Continued expansion in research on physical properties of magnetically ordered crystals and films requires exploring a wide spectrum of magnetic phenomena, including magnetoelectric ones. Such cross-correlated phenomena have attracted considerable attention in recent years because of novel physics and great potential for practical applications \cite{Fieb,Smol,ZP12}.

The polarization, induced by an external magnetic field and/or magnetic ordering in conventional magnetoelectric materials and multiferroics, is usually attributed to structural changes, e.g. displacements of sublattices. Along with the classic magnetoelectric mechanism caused by the direct interaction of the ferroelectric and magnetic order parameters \cite{Smol,ZP12,Turo}, the mechanism of inhomogeneous magnetoelectric interactions \cite{Barj,ZP09,Most} are also actively discussed. The interaction leads to new physical effects: the appearance of improper polarization in multiferroics, the flexomagnetoelectric surface effect \cite{ZP09}, and the electrical control of magnetic domain walls in films of iron garnets \cite{Logg}, etc.

It is well known that the space-time inversion symmetry breakdown is a necessary condition for the existence of the linear magnetoelectric effect in a material. Among many magnetoelectric materials are particularly interesting those with crystal structures which have a center of inversion (i.e. even with respect to the spatial inversion) and with magnetic structures which are odd with respect to the space and time. There exist quite a lot of such magnetic materials with the antiferromagnetic ordering, for example, the classical magnetoelectric Cr$_2$O$_3$.

In this context, very interesting are rare-earth orthoferrites ReFeO$_3$ in which there are two magnetic subsystems constituted by iron and rare-earth ions. The magnetic structure of the iron subsystem is even relative to the spatial inversion \cite{Yamg}, thus the linear magnetoelectric effect in it is forbidden by the symmetry. Indeed, in YFeO$_3$, LaFeO$_3$, and LuFeO$_3$, in which there is the only iron sublattice, the linear magnetoelectric effect is not observed. It is also not observed in other rare-earth orthoferrites at high temperatures when the rare-earth subsystem is disordered, being actually paramagnetic, and therefore spatially even. But, magnetically driven ferroelectricity was observed recently in electropoled orthochromites and is possible in isostructural orthoferrites as well \cite{Saha,Raje}.

The situation changes drastically when the temperature is lowered to the critical point of the rare-earth subsystem magnetic ordering and below it ($T_c \sim$ 1--5 K). In this case, the magnetic symmetry allows the existence of the linear magnetoelectric effect. The exchange interaction between the rare-earth and iron sublattices results in the crystal spontaneous electric polarization, i.e. at low temperatures these materials are multiferroics \cite{ZV08}. The phenomenon and other attendant effects were observed in DyFeO$_3$ \cite{DyFe}, GdFeO$_3$ \cite{GdFe}, SmFeO$_3$~\cite{SmFe}, etc. The key factor for such a behaviour is the presence of an antisymmetric (odd with respect to the space inversion) magnetic sublattice. From this point of view, very interesting are crystals of garnet structure with rare-earth sublattices.

A family of garnets provides a fascinating field of science and technology because of versatile functions that could be attained by introducing different ions \cite{Paol}. Studies of several years show \cite{Kric,Pisa,Pavl} that doping, creation of thin films, multilayer structures and composites based on yttrium iron garnet leads to strengthening a number of effects, as well as to the discovery of fundamentally new phenomena in these materials \cite{Yams,Koha}.

It is reported \cite{Yams} on the finding of the large magneto-capacitance effect in yttrium iron garnet, characterized by magnetically tunable quantum paraelectricity. The electric-dipole can be hosted by the impurity like Fe$^{2+}$ site, where the spin-orbit interaction governs the quantum-mechanical relaxation process in response to the external magnetic field or the magnetization vector.

Despite the significant value of the magnetoelectric effect in garnets \cite{Kric,Kaby}, these materials are non-traditional materials towards magnetoelectricity. Still unsolved remains the question on the nature of the first-order effect in yttrium iron garnet, observed \cite{Ogaw} by applying an electric field during cooling. Such emergence of the first-order magnetoelectric effect conflicts the crystal symmetry of yttrium iron garnet. Meanwhile, there are known several garnets of odd magnetic structures with respect to spatial inversion, for example, the most investigated dysprosium-aluminium garnet.

In this work, we investigate single-crystal garnets and garnet films as well by means of the symmetry and quantum mechanical combined analysis, which allows to reduce some of the uncertainties inherent in purely point symmetry analysis. We show that despite the antisymmetric magnetic structures in garnet crystals the linear magnetoelectric effect in most of them is not realized, therefore the condition of the magnetic subsystem antisymmetry is only a necessary condition for the effect to exist, not sufficient one. Nevertheless, in some garnets, for example, in Mn$_3$Al$_2$Si$_3$O$_{12}$, the existence of the linear magnetic effect is possible. We also show, that other new odd effects can be observed in garnets, namely, a piezoelectric effect induced by a magnetic field in Nd$_3$Ga$_5$O$_{12}$, a piezomagnetic effect in Dy$_3$Al$_5$O$_{12}$, etc.

We have also revealed an interesting feature of some garnets such as Dy$_3$Al$_5$O$_{12}$ or Ho$_x$Y$_{3-x}$Fe$_5$O$_{12}$ with rare-earth ions of Ising type. We show that magnetic fields induce electric dipole moments directly in the $4f$-electronic shell of a rare-earth ion and form in the materials fascinating antiferroelectric structures, though the polarization of the materials is zero on a macroscopic scale. But for all that, the magnetic structure of the garnets remains antisymmetrical. A strong enough magnetic field destroys the antisymmetric structures, converting them to ferromagnetic ones. In this case all of the above mentioned magnetoelectric effects disappear. This also applies to dysprosium iron garnet ($H_{\text{exch}} \sim 100$ kOe).

An interesting situation takes place in epitaxial films of mixed garnets R$^{(1)}_{3-x}$R$^{(2)}_x$M$_5$O$_{12}$, where R$^{(1)}$ and R$^{(2)}$ stand for different rare-earth ions and $M$ denotes Al or Fe. The uniform distribution of the rare-earth ions by their nonequivalent positions in the unit cell of the crystal in this case is disturbed, which results in the electric polarization occurrence in them. Another possible source for the linear magnetoelectric effect to arise in a bulk crystal is a spatial inhomogeneity of an external magnetic field or internal magnetic fields on domain boundaries. The latter inhomogeneity can be a source of magnetoelectricity for nanostructures and domain boundaries of the both Neel and even Bloch types, not requiring the Dzyaloshinskii-Moria mechanism.

The present paper is organized as follows. Section \ref{sec2} describes the structure of garnet crystals. The magnetoelectric Hamiltonian and the expression for the effective electric dipole moment operator of rare-earth ions in garnets are obtained in Section \ref{sec3}. Section \ref{sec4} deals with the symmetry analysis of the possible magnetic modes in garnets. On the basis of the analysis we discuss the possibility of the linear magnetoelectric and other odd effects existence in a single crystal (Section \ref{sec4}), describe the antiferroelectricity induced by magnetic field (Section \ref{sec5}), and demonstrate the possibility of the linear magnetoelectric effect in an inhomogeneous magnetic field in bulk crystals (Section \ref{sec6}). Finally, we depict the magnetoelectric effect in epitaxial garnet thin films owed to the non-equal occupancies of the ion $c$-positions in a primitive crystal cell by rare-earth ions (Section \ref{sec7}).


\section{Rare-earth garnets} \label{sec2}

Garnet crystals R$_3$M$_5$O$_{12}$, where R stands for a rare-earth or yttrium ion and M stands for a metal ion (such as Fe$^{3+}$, Al$^{3+}$, Ga$^{3+}$, etc.), have a complicated cubic crystallographic structure, described by the O$^{10}_h$ space group. While a primitive cell contains four R$_3$M$_5$O$_{12}$ units, a unit crystal cell contains eight of them.

Rare-earth ions are located in the positions with a dodecahedral environment of oxygen ions with no inversion center (the so-called $c$-positions of the D$_2$ symmetry). Obviously, there are twelve $c$-positions in each primitive cell, differing from each other by symmetry axis orientations. The coordinates of the non-equivalent $c$-positions (expressed in the units of the cell edge lengths) and their symmetry axes are given in Table I. We should note, that positions 7--12 are connected with positions 1--6 by the $I$ and $C_2$ operations, where symbols $I$ and $C_2$ stand for the inversion and the rotation by $180^{\circ}$ about a local axis, relatively. In so doing, ${\bf e}^{(k+6)}_{x,z} = {\bf e}^{(k)}_{x,z}$ and ${\bf e}^{(k+6)}_{y} = - {\bf e}^{(k)}_{y}$, i.e. ${\bf e}^{(7)} = C_2(y)I \cdot {\bf e}^{(1)}$.

\begin{table} \caption{\label{tab1} The coordinates and the symmetry axes of the $c$-positions}
\begin{ruledtabular} \begin{tabular}{ccccccccccccc}
    $k$ & 1 & 2 & 3 & 4 & 5 & 6 & 7 & 8 & 9 & 10 & 11 & 12 \\
	\hline
	${\bf r}^{(k)}$ &
	$0 \: \frac{3}{4} \: \frac{3}{8}$ & $0 \: \frac{1}{4} \: \frac{1}{8}$ & $\frac{3}{8} \: 0 \: \frac{3}{4}$ & $\frac{1}{8} \: 0 \: \frac{1}{4}$ &
	$\frac{3}{4} \: \frac{3}{8} \: 0$ & $\frac{1}{4} \: \frac{1}{8} \: 0$ & $0 \: \frac{1}{4} \: \frac{5}{8}$ & $0 \: \frac{3}{4} \: \frac{7}{8}$ &
	$\frac{5}{8} \: 0 \: \frac{1}{4}$ & $\frac{7}{8} \: 0 \: \frac{3}{4}$ & $\frac{1}{4} \: \frac{5}{8} \: 0$ & $\frac{3}{4} \: \frac{7}{8} \: 0$ \\
	${\bf e}^{(k)}_x$ & 110 & $1\bar{1}0$ & 011 & $01\bar{1}$ & 101 & $\bar{1}01$ & 110 & $1\bar{1}0$ & 011 & $01\bar{1}$ & 101 & $\bar{1}01$ \\
	${\bf e}^{(k)}_y$ & $\bar{1}10$ & 110 & $0\bar{1}1$ & 011 & $10\bar{1}$ & 101 & $1\bar{1}0$ & $\bar{1}\bar{1}0$ & $01\bar{1}$ & $0\bar{1}\bar{1}$ &
    $\bar{1}01$ & $\bar{1}0\bar{1}$ \\
	${\bf e}^{(k)}_z$ & 001 & 001 & 100 & 100 & 010 & 010 & 001 & 001 & 100 & 100 & 010 & 010 \\
\end{tabular} \end{ruledtabular} \end{table}

There is no space inversion $I$ the symmetry group of the rare-earth ion environment, therefore the crystal field operator of rare-earth ions includes not only $I$-even terms, but also odd terms. The operator can be expressed as
    \begin{equation} \label{crys} 
        {\cal H}_{\text{cr}} = {\cal H}^{\text{even}}_{\text{cr}} + {\cal H}^{\text{odd}}_{\text{cr}},
    \end{equation}
where
    $$
        {\cal H}^{\text{even}}_{\text{cr}} = B^2_0 {\cal C}^2_0 + B^4_0 {\cal C}^4_0 + B^6_0 {\cal C}^6_0
        + B^2_2 \left( {\cal C}^2_{-2} + {\cal C}^2_2 \right) + B^4_2 \left( {\cal C}^4_{-2} + {\cal C}^4_2 \right)
        + B^6_2 \left( {\cal C}^6_{-2} + {\cal C}^6_2 \right)
    $$
and
    $$
		{\cal H}^{\text{odd}}_{\text{cr}} = i B^3_2 \left( {\cal C}^3_{-2} - {\cal C}^3_2 \right)
        + i B^5_2 \left( {\cal C}^5_{-2} - {\cal C}^5_2 \right) + i B^5_4 \left( {\cal C}^5_{-4} - {\cal C}^5_4 \right).
	$$
Here $B^p_q$ are the crystal field parameters and ${\cal C}^p_q = \sum_{k=1}^n {\cal C}^p_q (k)$, where $n$ is the number of electrons in the $4f$-shell and ${\cal C}^p_q (k)$ are the single-electron irreducible tensor operators, which are defined by the reduced matrix elements $\langle l' || {\cal C}^p || l \rangle = \sqrt{2l+1} \, C^{l'0}_{l0p0}$. Here $C^{lm}_{l_1m_1l_2m_2}$ are the Clebsch-Gordan coefficients.

The odd crystal field mixes the states of the ground $4f^n$ configuration with the states of the $4f^{n-1}5d^1$ and $4f^{n-1}4g^1$ configurations, which are of different evenness compared with $4f^n$ states. This results in a possibility to induce the electric dipole moment in the rare-earth ions by a magnetic field. In this work, we calculate the induced polarization in terms of an ion response to external electric field ${\bf E}$.


\section{The magnetoelectric Hamiltonian} \label{sec3}

The actual perturbation Hamiltonian $\cal V$ of a rare-earth ion in an external electric field reads
    \begin{equation} \label{pert} 
		{\cal V} = - {\bf d} {\bf E} + {\cal H}^{odd}_{cr},
	\end{equation}
where ${\bf E}$ is the strength of the external electric field and ${\bf d} = -e \sum^{n}_{k=1} {\bf r}_k$ is the dipole moment operator of the ion with $n$ electrons in the unfilled $4f$ shell.

The linear on the strength of the applied electric field corrections to the ion energy levels arises in the second-order perturbation theory with small parameter $|| {\cal V} || / W$, where $|| {\cal V} ||$ is the norm of the $\cal V$ operator and $W$ is the energy difference between the ground states and the weight center of excited ion electronic configurations (typically, $W \sim 10^5$ cm$^{-1}$ for rare-earth ions). For the sake of simplicity, we will take into account only the lowest excited $4f^{n-1} 5d^1$ configuration.

Making use of the wave-function genealogical scheme construction and the quantum theory of angular momentum \cite{VMCH}, we derived the expression for the magnetoelectric operator of garnets. The details of the calculations are given in ref. \cite{2013}. The final expression for the magnetoelectric Hamiltonian is
    \begin{equation} \label{hmee} 
        {\cal H}_{me} = - \left( \frac{e r_{fd}}{W} \right) \cdot {\bf E D} =
        - \left( \frac{e r_{fd}}{W} \right) \cdot \left( E_{+} D_{-} + E_{-} D_{+} + E_z D_z \right),
    \end{equation}
where $r_{fd}$ is the radial integral and $D_{\alpha}$ ($\alpha = x, y, z$) are the effective operators of the rare-earth ion electric dipole moment components. For the sake of brevity, we use the notation of so called cyclic operators $E_{\pm} = \left( E_x \pm i E_y \right) / \sqrt{2}$ and $D_{\pm} = \left( D_x \pm i D_y \right) / \sqrt{2}$, where
    \begin{eqnarray} \label{Dpmz} 
        D_{\pm} & = & 2i \left( \sum_{p=2,4,6} b^p_1 {\cal C}^p_{\mp 1} + \sum_{p=4,6} \left( b^p_{31} {\cal C}^p_{\pm 3} + b^p_{32} {\cal C}^p_{\mp 3} \right) + b^6_5 {\cal C}^6_{\pm 5} \right), \nonumber \\
        D_z     & = & 2i \left( \sum_{p=2,4,6} b^p_2 \left( {\cal C}^p_{-2} - {\cal C}^p_2 \right) +
        \sum_{p=4,6} b^p_4 \left( {\cal C}^p_{-4} - {\cal C}^p_4\right) \right)
    \end{eqnarray}
Coefficients $b^p_q$ in eq. \eqref{Dpmz} can be expressed in the terms of the crystal field parameters,
    \begin{equation*} \begin{matrix}
        \textstyle b^2_1 = \frac{2\sqrt{2}}{7} B^3_2, \text{ } b^4_1 = \frac{\sqrt{15}}{7} B^5_2 - \frac{11}{14\sqrt{42}} B^3_2, \text{ } b^6_1 =
        - \frac{\sqrt{65}}{7\sqrt{22}} B^5_2, \\
        \textstyle b^4_{31} = \frac{11}{14\sqrt{6}} B^3_2 - \frac{\sqrt{15}}{7\sqrt{7}} B^5_2, \text{ } b^4_{32} = \frac{6\sqrt{5}}{7\sqrt{7}} B^5_4, \text{ } b^6_{31} = \frac{3\sqrt{13}}{7\sqrt{11}} B^5_2, \text{ } b^6_{32} = - \frac{\sqrt{39}}{14\sqrt{11}} B^5_4, \\
        \textstyle b^6_5 = \frac{\sqrt{65}}{14} B^5_4, \text{ } b^2_2 = \frac{2}{7\sqrt{7}} B^3_2, \text{ } b^4_2 = \frac{11}{7\sqrt{42}} B^3_2 + \frac{\sqrt{15}}{7} B^5_2, \text{ } b^6_2 = \frac{2\sqrt{26}}{7\sqrt{11}} B^5_2 \\
        \textstyle b^4_4 = \frac{3\sqrt{5}}{7\sqrt{7}} B^5_4, \text{ } b^6_4 = \frac{\sqrt{65}}{7\sqrt{11}} B^5_4.
    \end{matrix} \end{equation*}
In the second-order approximation, the magnetoelectric Hamiltonian in eq. \eqref{hmee} reads
    $$ {\cal H}^{(2)}_{me} = - \left( \frac{e r_{fd}}{W} \right) \cdot \left( \sqrt{2} i E_x b^2_1 \left( {\cal C}^2_{-1} + {\cal C}^2_1 \right) +
    \sqrt{2} E_y b^2_1 \left( {\cal C}^2_{-1} - {\cal C}^2_1 \right) + i E_z b^2_2 \left( {\cal C}^2_{-2} - {\cal C}^2_2 \right) \right). $$

Now we have to average the obtained expression for the magnetoelectric Hamiltonian in eq.~\eqref{hmee} with the equilibrium density matrix defined by the non-perturbed Hamiltonian
    \begin{equation}
		{\cal H}_0 = {\cal H}_z + {\cal H}^{\text{even}}_{\text{cr}}, \nonumber
	\end{equation}	
where ${\cal H}_z$ is the Hamiltonian of interionic exchange interactions and interactions of the ions with an external magnetic field. The crystal field Hamiltonian ${\cal H}^{\text{even}}_{\text{cr}}$ is defined in eq.~\eqref{crys}. As a result, we obtain the operator of the magnetoelectric interaction per one primitive cell,
    \begin{equation} \label{cell} 
        {\cal H}_{me} = - \left( \frac{e r_{fd}}{W} \right) {\bf E} \sum_{k=1}^{12} \langle {\bf D}_k \rangle,
    \end{equation}
where ${\bf D}_k$ is the effective dipole moment of an ion located in the $k$-th $c$-position. Making use of eq. \eqref{cell}, we find the expressions for the effective dipole moment components of the $k$-th ion (in the local axes),
    \begin{equation} \label{polk} 
		P^{(k)}_{\alpha} = - \frac{\partial E_{me}}{\partial E^{(k)}_{\alpha}} =
        - \left( \frac{e r_{fd}}{W} \right) \langle D_{k \alpha} \rangle.
	\end{equation}

As it follows from eq.~\eqref{Dpmz}, $\langle D_{k \alpha} \rangle = 0$ and $P^{(k)}_{\alpha} = 0$ if a magnetic field or a magnetic ordering are not present, because the eigenfunctions of the ${\cal H}^{\text{even}}_{\text{cr}}$ operator at $H = 0$ are the set of spherical harmonics different from each other in the magnetic quantum number by $\pm 2$ and $\langle {\cal C}_{-2} \rangle = \langle {\cal C}_2 \rangle$.

The role of the ${\cal H}^{\text{even}}_{\text{cr}}$ crystal field is to split the multiplets of rare-earth ions with $L \ne 0$ into doublets, quasi-doublets, or singlets. If the doublets and quasi-doublets are of Kramers type, then
    \begin{equation} \label{dalp} 
        \langle D_{\alpha} \rangle = \sum_{ij} C_{\alpha i j} H_i M_j
    \end{equation}
in the presence of a magnetic field at low temperatures \cite{2013,Book,Babu,Nekv}. $C_{\alpha i j}$ are certain numerical coefficients, determined by the wave functions and the energy levels of a rare-earth ion in the crystal field. The coefficients are of order of several unities \cite{2013}. Components $M_j(H,T)$ of the mean ion magnetic moment are owed to the splitting of the ground doublet (or quasi-doublet) levels in the magnetic field. In this case the dipole moment linearly depends on the strong field ($\mu J H > k T$), in which the ion magnetic moment is nearly saturated. The ion effective dipole moment is then estimated to be $10^{-22} \text{ esu} \cdot \text{cm}$ per kOe.

As for singlets, $$ \langle D_{\alpha} \rangle = \sum_{ij} q_{\alpha i j} H_i H_j. $$ Here and in eq.~\eqref{dalp} as well, $H$ is an effective magnetic field, comprising an external $H_0$ and an exchange (dipole) $H_{\text{exch}}$ magnetic fields.


\section{Linear magnetoelectric and other odd effects} \label{sec4}

The most interesting are garnet crystals with rare-earth ions of doublet or quasi-doublet ground states. In order to determine the electric dipole structure of the rare-earth ions in the garnets it is convenient to use symmetry analysis. The contributions of a single rare-earth ion into the magnetoelectric energy of the crystal are the invariants under the transformations of the $D_2$ group. These invariants are composed from vectors ${\bf E}$, ${\bf H}$, and ${\bf M}$, see eq.~\eqref{cell} and eq.~\eqref{dalp}. In the general case, the invariant combination reads (in local axes of the $k$-th position),
    \begin{eqnarray} \label{eq08} 
        E^{(k)}_{me} & = & C_1 E^{(k)}_x H^{(k)}_y M^{(k)}_z + C_2 E^{(k)}_y H^{(k)}_x M^{(k)}_z + C_3 E^{(k)}_x H^{(k)}_z M^{(k)}_y \nonumber \\
                     & + & C_4 E^{(k)}_z H^{(k)}_x M^{(k)}_y + C_5 E^{(k)}_y H^{(k)}_z M^{(k)}_x + C_6 E^{(k)}_z H^{(k)}_y M^{(k)}_x.
    \end{eqnarray}

As it follows from eq.~\eqref{eq08}, the configuration of the rare-earth ion electric dipole moments is determined by the orientation of the magnetic field and the magnetic structure of the garnet crystal, which is meant as distribution of the rare-earth ion magnetic moments over the twelve $c$-positions in a primitive cell.

The existence of a linear magnetoelectric effect (then $\sum_k E^{(k)}_{me} \ne 0$) is possible only for the $I$-odd magnetic structures, for which ${\bf M}^{(k+6)} = - {\bf M}^{(k)}$. This obviously follows from eq.~\eqref{eq08} if one takes into account that ${\bf e}^{(k+6)}_{x,z} = {\bf e}^{(k)}_{x,z}$ and ${\bf e}^{(k+6)}_{y} = - {\bf e}^{(k)}_{y}$, see Table \ref{tab1}. The mentioned requirement for the magnetic structure of rare-earth ions is a necessary condition, but, as it will be shown below, is not sufficient one.

The magnetic moments of rare-earth ions form magnetic structures, which can be described by the certain modes. The magnetic modes of rare-earth crystals with the garnet structure were described in ref. \cite{ZV87} using the notation of the Kovalev's handbook \cite{Kova}. To construct the modes it is enough to know the irreducible representations of the $O_h$ point group (denoted as $\tau_{\nu}$) and the non-trivial (accompanying) translations of the $O^{10}_h$ space group. The $O_h$ group includes four one-dimensional ($\nu = 1, 2, 3, 4$), two two-dimensional ($\nu = 5, 6$) and four three-dimensional ($\nu = 7, 8, 9, 10$) irreducible representations. It was shown in ref. \cite{ZV87} that the magnetic representation is reducible and has the following composition: $n_{1,2} = 0$, $n_{3,4,5,6} = 1$, $n_{7,8} = 2$, and $n_{9,10} = 3$.

As a magnetic mode of the rare-earth subsystem we will mean below certain linear combinations of the magnetic moment components of the twelve rare-earth ions, being transformed according to the $\tau_{\nu}$ irreducible representations of the $O_h$ group. The magnetic modes will be denoted as $\eta^{\nu,\mu}_{\lambda}$, where index $\lambda$ stands for a row of the $\tau_{\nu}$ representation with the dimension greater than one ($\mu = 1,2$ for $\tau_{7,8}$ and $\mu = 1,2,3$ for $\tau_{9,10}$). The $\eta^{\nu,\mu}_{\lambda}$ modes can be expressed as
    \begin{equation} \label{grek} 
        \eta^{\nu,\mu}_{\lambda} = \eta^{\nu,\mu}_{\lambda} \left( M^{(k)} \right) - (-1)^{\nu} \eta^{\nu,\mu}_{\lambda} \left( M^{(k+6)} \right),
    \end{equation}
where term $\eta^{\nu,\mu}_{\lambda} \left( M^{(k)} \right)$ represents the contribution from the magnetic moments of the first six ions ($k = 1, \ldots, 6$). The contribution from the other six ions ($k = 7, \ldots, 12$) is determined by the second term in eq.~\eqref{grek}. The $\eta^{\nu,\mu}_{\lambda} \left( M^{(k)} \right)$ modes are given in Table \ref{tab2} (the magnetic moment components of the rare-earth ions are given in the crystallographic coordinate system). The $I$-even magnetic structures are described by the magnetic modes with odd $\nu$ and vice versa.

\begin{table} \caption{\label{tab2} The magnetic modes of the rare-earth subsystem}
\begin{ruledtabular} \begin{tabular}{cc}
$\tau_{\nu}$        & $\eta^{\nu,\mu}_{\lambda} (M^{(k)})$, \;\;\; $k = 1, \ldots, 6$ \\
\hline
$\tau_3$, $\tau_4$  & $\eta^{\nu} (M^{(k)}) = M_x^{(4)} - M_x^{(3)} + M_y^{(6)} - M_y^{(5)} + M_z^{(2)} - M_z^{(1)}$ \;\;\; ($\nu = 3, 4$) \\
                    & $\tau_3 : H_x H_y H_z$; \;\;\; $\tau_4 : E_x H_y H_z + E_y H_x H_z + E_z H_x H_y$; \\
                    & $H_x E_x (\varepsilon_{yy} - \varepsilon_{zz}) + H_y E_y (\varepsilon_{zz} - \varepsilon_{xx}) + H_z E_z (\varepsilon_{xx} - \varepsilon_{yy})$ \\
\hline
$\tau_5$, $\tau_6$  & $\eta^{\nu}_1 (M^{(k)}) = \frac{1}{2} \left( M_x^{(4)} - M_x^{(3)} - M_y^{(6)} + M_y^{(5)} \right)$ \;\;\; ($\nu = 5, 6$) \\
                    & $\eta^{\nu}_2 (M^{(k)}) = \frac{\sqrt{3}}{2} \left( M_z^{(2)} - M_z^{(1)} - \frac{1}{3} \eta^{\nu-2} (M^{(k)}) \right)$ \\
                    & $\varepsilon^5_1 = \frac{1}{2} (\varepsilon_{xx} - \varepsilon_{yy})$, \;\;\; $\varepsilon^5_2 = \frac{\sqrt{3}}{2} \left( \varepsilon_{zz} - \frac{1}{3} (\varepsilon_{xx} + \varepsilon_{yy} + \varepsilon_{zz}) \right)$ \\
\hline
$\tau_7$, $\tau_8$  & $\eta^{\nu,1}_1 (M^{(k)}) = M_x^{(1)} + M_x^{(2)} - M_x^{(5)} - M_x^{(6)}$, \\
                    & $\eta^{\nu,1}_2 (M^{(k)}) = - M_y^{(1)} - M_y^{(2)} + M_y^{(3)} + M_y^{(4)}$, \\
                    & $\eta^{\nu,1}_3 (M^{(k)}) = - M_z^{(3)} - M_z^{(4)} + M_z^{(5)} + M_z^{(6)}$, \\
                    & $\eta^{\nu,2}_1 (M^{(k)}) = - M_y^{(1)} + M_y^{(2)} + M_z^{(5)} - M_z^{(6)}$, \\
                    & $\eta^{\nu,2}_2 (M^{(k)}) = M_x^{(1)} - M_x^{(2)} - M_z^{(3)} + M_z^{(4)}$, \\
                    & $\eta^{\nu,2}_3 (M^{(k)}) = M_y^{(3)} - M_y^{(4)} - M_x^{(5)} + M_x^{(6)}$, \;\;\; $(\nu = 7, 8)$ \\
\hline
$\tau_9$, $\tau_{10}$ & $\eta^{\nu,1}_1 (M^{(k)}) = M_x^{(3)} + M_x^{(4)}$, \;\;\;
                        $\eta^{\nu,1}_2 (M^{(k)}) = M_y^{(5)} + M_y^{(6)}$, \;\;\;
                        $\eta^{\nu,1}_3 (M^{(k)}) = M_z^{(1)} + M_z^{(2)}$, \\
                      & $\eta^{\nu,2}_1 (M^{(k)}) = M_x^{(1)} + M_x^{(2)} + M_x^{(5)} + M_x^{(6)}$, \\
                      & $\eta^{\nu,2}_2 (M^{(k)}) = M_y^{(1)} + M_y^{(2)} + M_y^{(3)} + M_y^{(4)}$, \\
                      & $\eta^{\nu,2}_3 (M^{(k)}) = M_z^{(3)} + M_z^{(4)} + M_z^{(5)} + M_z^{(6)}$, \\
                      & $\eta^{\nu,3}_1 (M^{(k)}) = -M_y^{(1)} + M_y^{(2)} - M_z^{(5)} + M_z^{(6)}$, \\
                      & $\eta^{\nu,3}_2 (M^{(k)}) = -M_x^{(1)} + M_x^{(2)} - M_z^{(3)} + M_z^{(4)}$, \\
                      & $\eta^{\nu,3}_3 (M^{(k)}) = -M_y^{(3)} + M_y^{(4)} - M_x^{(5)} + M_x^{(6)}$, \;\;\; $(\nu = 9, 10)$ \\
                      & $\varphi^9_1 = E_x (K_y H_y + K_z H_z)$, \;\;\;
                        $\varphi^9_2 = E_y (K_x H_x + K_z H_z)$, \;\;\;
                        $\varphi^9_3 = E_z (K_x H_x + K_y H_y)$, \\
                      & $\varphi^{10}_1 = \left[ {\bf E} \times {\bf H} \right]_x$, \;\;\;
                        $\varphi^{10}_2 = \left[ {\bf E} \times {\bf H} \right]_y$, \;\;\;
                        $\varphi^{10}_3 = \left[ {\bf E} \times {\bf H} \right]_z$ \\
\end{tabular} \end{ruledtabular} \end{table}

As above mentioned, the linear magnetoelectric effect in rare-earth garnets can exist only in the $I$-odd magnetic structures. The magnetoelectric energy of a crystal $\sum_{k=1}^{12} E_{\text{me}}^{(k)}$, where $E_{\text{me}}^{(k)}$ are given by eq.~\eqref{eq08}, is the invariant (relative to the transformations of the $O^{10}_h$ symmetry group, which includes only non-trivial translations) combination constructed from some products of the odd $\eta^{\nu,\mu}_{\lambda}$ magnetic modes components and basis functions $\varphi^{\nu,\mu}_{\lambda} \left( \{ E, H \} \right)$, which are bilinear combinations over $E$ and $H$.

It is known only two odd magnetic structures existing in garnets. One of them is described by the $\eta^4 (M)$ mode, which corresponds to the $\tau_4$ irreducible representation (see Table \ref{tab2}), and exists in Nd$_3$Ga$_5$O$_{12}$ \cite{Izyu} at low temperatures $T < T^{*}$. In this structure
    \begin{eqnarray*}
        {\bf M}^{(4)} = -{\bf M}^{(3)} = -{\bf M}^{(10)} = {\bf M}^{( 9)} = [100], \\
        {\bf M}^{(6)} = -{\bf M}^{(5)} = -{\bf M}^{(12)} = {\bf M}^{(11)} = [010], \\
        {\bf M}^{(2)} = -{\bf M}^{(1)} = -{\bf M}^{( 8)} = {\bf M}^{( 7)} = [001].
    \end{eqnarray*}
Nevertheless, there is no linear magnetoelectric effects in such a structure, as shows the direct summing over the $k$-index in eq.~\eqref{eq08}. The reason is that there are no bilinear over $E$ and $H$ functions transformed as the $\tau_4$ representation. At the same time, the given structure admits the invariant $$ I = \left[ E_x H_x \left( \varepsilon_{yy} - \varepsilon_{zz} \right) + E_y H_y \left( \varepsilon_{zz} - \varepsilon_{xx} \right) + E_z H_z \left( \varepsilon_{xx} - \varepsilon_{yy} \right) \right] \eta^4, $$ where $\varepsilon_{\alpha \beta}$ are the components of the deformation tensor. The invariant describes the appearance of a magnetic field induced piezoelectric effect in the $\tau_4$ antiferromagnetic phase realized in Nd$_3$Ga$_5$O$_{12}$ garnet. The effect is magnetic field odd.

The other odd magnetic structure exists at low temperatures in Mn$_3$Al$_2$Si$_3$O$_{12}$ \cite{Izyu},
    \begin{eqnarray*}
        {\bf M}^{(4)} = {\bf M}^{(3)} = - {\bf M}^{(10)} = - {\bf M}^{( 9)} = [\bar{u} \, v \, v], \\
        {\bf M}^{(6)} = {\bf M}^{(5)} = - {\bf M}^{(12)} = - {\bf M}^{(11)} = [v \, \bar{u} \, v], \\
        {\bf M}^{(2)} = {\bf M}^{(1)} = - {\bf M}^{( 8)} = - {\bf M}^{( 7)} = [v \, v \, \bar{u}],
    \end{eqnarray*}
where $v = 1 - \sqrt{3}$ and $u = 2 - \sqrt{3}$. This magnetic structure is transformed according to the $\tau_{10}$ representation and described by the set of the $\eta^{10,1}_{\lambda}$ and $\eta^{10,2}_{\lambda}$ modes, see Table~\ref{tab2}. $\eta^{10,1}_{\lambda} = -2u$ and $\eta^{10,2}_{\lambda} = 4v$.

The magnetoelectric energy of the crystal for the structure differs from zero and can be made up of the two following invariants,
    \begin{equation}
        I_{\mu} = \sum_{\lambda = 1}^{3} \eta^{10,\mu}_{\lambda} \varphi^{10}_{\lambda} (E, H),
    \end{equation}
where $\mu = 1,2$. The $\varphi^{10}_{\lambda}$ quantities are given in Table~\ref{tab2}, $\varphi^{10}_{1} = \left[ {\bf E} \times {\bf H} \right]_x$, $\varphi^{10}_{2} = \left[ {\bf E} \times {\bf H} \right]_y$, and $\varphi^{10}_{3} = \left[ {\bf E} \times {\bf H} \right]_z$. The bilinearity of the invariants indicates the possibility of a linear magnetoelectric effect in Mn$_3$Al$_2$Si$_3$O$_{12}$ garnet at low temperatures.


\section{Antiferroelectricity induced by magnetic field} \label{sec5}

As already mentioned, a linear magnetoelectric effect does not occur in most of garnets because of spatial evenness. However, a magnetic field can make up various antiferroelectric structures in garnet crystals. The most interesting to study is the electric dipole moments induced by magnetic fields in strongly anisotropic rare-earth Ising ions of rare-earth iron garnets, for example, Sm$^{3+}$ ions in Sm$_3$Fe$_5$O$_{12}$ and Ho$^{3+}$ ions in Ho$_x$Y$_{3-x}$Fe$_5$O$_{12}$.

We start our consideration with dysprosium aluminium garnet Dy$_3$Al$_5$O$_{12}$, which is the most investigated cubic magnet with Ising ions. The ground state of a dysprosium ion in the crystal field is the Kramers doublet, separated from excited states by the energy of roughly 100 K. The $g$-tensor components of this state in the local coordinate system are $g_z \approx 18$ and $g_x \sim g_y < 0{,}5$ \cite{Wolf}. Such strong anisotropy makes possible the dysprosium ions be treated as Ising ions with magnetization axes coinciding with the $z_k$ local axes given in Table \ref{tab1}.

The exchange interaction between the dysprosium ions in the dysprosium aluminium garnet is antiferromagnetic. This compound is a multi-sublattice antiferromagnet with the $T_N = 2.54$ K ordering temperature. Below the Neel point the magnetic structure of the crystal in magnetic field ${\bf H} || [111]$ can be formed by two different configurations of the magnetic moments,
    \begin{eqnarray} \label{ord1}
		{\bf M}_2 = - {\bf M}_1 = - {\bf M}_7 = {\bf M}_8 = [001], \nonumber \\
		{\bf M}_4 = - {\bf M}_3 = - {\bf M}_9 = {\bf M}_{10} = [100], \\
		{\bf M}_6 = - {\bf M}_5 = - {\bf M}_{11} = {\bf M}_{12} = [010], \nonumber
	\end{eqnarray}
and
    \begin{eqnarray} \label{ord2}
		{\bf M}_2 = {\bf M}_1 = {\bf M}_7 = {\bf M}_8 = [001], \nonumber \\
		{\bf M}_4 = {\bf M}_3 = {\bf M}_9 = {\bf M}_{10} = [100], \\
		{\bf M}_6 = {\bf M}_5 = {\bf M}_{11} = {\bf M}_{12} = [010]. \nonumber
	\end{eqnarray}

The magnetic structure in eq.~\eqref{ord1} is described by the $\eta^3 (M)$ mode, which corresponds to the $\tau_3$ irreducible representation (see Table \ref{tab2}). The $\eta^3 (M)$ coefficient is the antiferromagnetic order parameter.

The structure determined by eq.~\eqref{ord2} is described in its turn by the irreducible representation $\tau^1_9$ with the $\eta^{9,1}_{\lambda}$ modes, see Table~\ref{tab2},
    \begin{eqnarray}
        \eta^{9,1}_1 (M) & = & M^4_x + M^3_x + M^9_x + M^{10}_x, \nonumber \\
        \eta^{9,1}_2 (M) & = & M^6_y + M^5_y + M^{11}_y + M^{12}_y, \nonumber \\
        \eta^{9,1}_3 (M) & = & M^2_z + M^1_z + M^8_z + M^7_z. \nonumber
    \end{eqnarray}
The $\eta^{9,1}_{\lambda}$ coefficients are the ferromagnetic order parameter, usually denoted as $M$.

Below the tricritical temperature ($T < T_k = 1{,}66$ K), parameters $\eta^{9,1}_{\lambda} \approx 0$ if $H < H_0$ and $\eta^3 \approx 0$ if $H > H_0$. Here $H_0 \approx 4$ kOe stands for the metamagnetic transition field. The interaction of the antiferromagnetic order parameter with an external magnetic fields is described by invariant $H_S \cdot \eta^3$, where $H_S$ is the induced staggered field ($H_S = H_x, H_y, H_z$). This interaction results in the energy difference of the two antiferromagnetic phases, namely, $A^{+}$ with $\eta^3 > 0$ and $A^{-}$ with $\eta^3 < 0$. The $A^{+}$ phase is stabilized by external field ${\bf H} || [111]$ at $T = 1.35$ K \cite{Gior}. The transition from the $A^{+}$ phase to the $A^{-}$ phase is likely to occur while the temperature decreases below 1.3 K \cite{Kolm}.

Besides the interaction with the induced staggered field, the antiferromagnetic order parameter $\eta^3$ also interacts with a crystal deformation in the presence of a magnetic field. This interaction is described by the invariant
    \begin{equation*}
        I(\varepsilon, H) = A \eta^3 (\varepsilon_{xy} H_z + \varepsilon_{xz} H_y + \varepsilon_{yz} H_x),
    \end{equation*}
where $A$ is the coupling constant and $\varepsilon_{\alpha \beta}$ are the components of the deformation tensor. The existence of such an invariant results in a piezomagnetic effect in the considered Dy$_3$Al$_5$O$_{12}$ garnet crystal, namely, in the initiation of a magnetization in the antiferromagnetic phase owed to the deformation,
    \begin{equation*}
        M_x = - A \eta^3 \varepsilon_{yz}, \;\;\; M_y = - A \eta^3 \varepsilon_{xz}, \;\;\; M_z = - A \eta^3 \varepsilon_{xy}.
    \end{equation*}
The magnetization can be detected in a magnetooptical experiment, for example, by means of the Faraday effect. Of course, there can also exist the inverse effect, the occurrence of a linear in field $H$ and odd deformation $\varepsilon \sim \eta^3 H$ in the antiferromagnetic phase. This phenomenon resembles the behavior of the features discovered by Dillon et al. \cite{Dill} in the magnetic linear birefringence of Dy$_3$Al$_5$O$_{12}$ at low temperatures.

\begin{figure} \includegraphics[scale=1.3192]{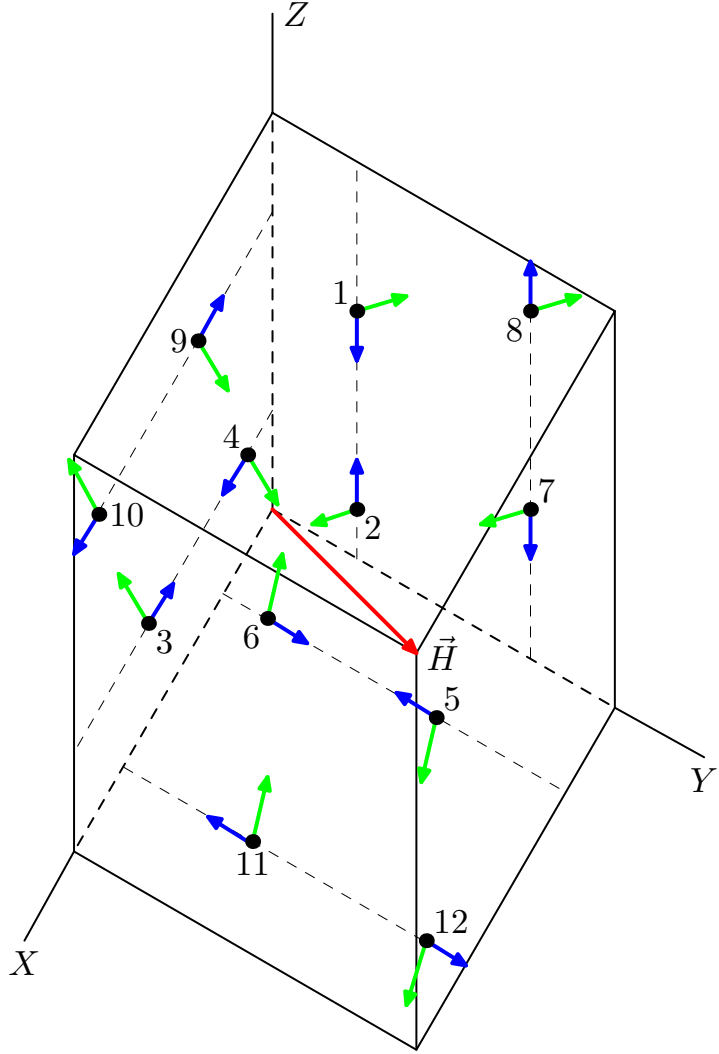} \caption{\label{afes} The magnetic structure of dysprosium aluminium garnet Dy$_3$Al$_5$O$_{12}$ \cite{Hast} and the corresponding structure of the electric dipole moments. The red arrow shows the direction of the external magnetic field. The blue arrows stand for the dysprosium ion magnetic moments and the green ones stand for the electric dipole moments induced by the external magnetic field.} \end{figure}

Now, we proceed to the discussion of the electric dipole moments in Dy$_3$Al$_5$O$_{12}$. Making use of eq.~\eqref{eq08} we find
    \begin{equation} \label{eq13} 
        P^{(k)}_x = -C_1 H^k_y M^k_z, \;\;\; P^{(k)}_y = -C_2 H^k_x M^k_z, \;\;\; P^{(k)}_z = 0.
    \end{equation}
For the stabilized $A^{+}$ phase in the field ${\bf H} || [111]$,
    \begin{equation*}
		{\bf P}^{(k)} = \sqrt{2/3} C_2 H \mu {\bf e}^{(k)}_{y} \;\;\; \text{ and } \;\;\; {\bf P}^{(k)} = \sqrt{2/3} C_1 H \mu {\bf e}^{(k)}_{x},
	\end{equation*}
for odd and even $k$, relatively. Here $\mu$ stands for the magnetic moment of a dysprosium ion, $\mu = 10 \mu_B$. The magnetic structure and the concomitant antiferroelectric structure for the $A^{+}$ magnetic phase ($\eta^3 > 0$) are shown in fig.~\ref{afes}.

Another example of a cubic magnet with Ising ions is holmium-yttrium iron garnet Ho$_x$Y$_{3-x}$Fe$_5$O$_{12}$, where $0 < x \leqslant 3$. As shown in ref.~\cite{Nova}, the ground state of a holmium ion in Ho$_3$Ga$_5$O$_{12}$ and Ho$_3$Al$_5$O$_{12}$ is a quasi-doublet, well separated from excited states. The doublet responds only to the external magnetic field projection on the local $z$-axis. Under the assumption that the same spectrum of holmium ions is realized in iron garnets, it is possible to explain quite a number of unique properties of holmium-yttrium iron garnets, in particular, the peculiarities of the field-induced orientational phase transitions \cite{Babu}, the peculiarities of resonance properties \cite{Levi,Plis}, etc.

It is easy to show, that the $f$-$d$ exchange interaction can induce an antiferroelectric structure in Ho$_x$Y$_{3-x}$Fe$_5$O$_{12}$. Rare-earth ions in the iron garnet are magnetized by effective magnetic field, $$ {\bf H}_{\text{eff}} = {\bf H} - \lambda {\bf M}_d, $$ therefore the structure of the holmium ion magnetic moments in holmium-yttrium iron garnets takes the form (at $T = 0$ K)
    \begin{equation} \label{magk} 
		{\bf M}^{(k)} = \mu_z {\bf e}^{(k)}_{z} \sign \left( {\bf H}_{\text{eff}} {\bf e}^{(k)}_{z} \right),
	\end{equation}
where $\mu_z \sim 10 \mu_B$ is the matrix element value of the magnetic moment operator in the quasi-doublet state of the holmium ion \cite{Wolf}. The components of the holmium ion electric dipole moments ${\bf P}^{(k)}$ in Ho$_x$Y$_{3-x}$Fe$_5$O$_{12}$ are given by eq.~\eqref{eq13} with $M^{(k)}_{z}$ defined by eq.~\eqref{magk}. Because positions 7--12 are related with positions 1--6 by the inversion and rotation operations, ${\bf P}^{(k+6)} = - {\bf P}^{(k)}$, $k = 1, \ldots, 6$ and thus ${\bf P} = \sum_k {\bf P}_k = 0$.


\section{The linear magnetoelectric effect in a spatially inhomogeneous magnetic field} \label{sec6}

The polarization in the above mentioned highly anisotropic mono-crystal garnets is possible at low temperatures in the case of an inhomogeneous magnetic field. We consider the following simple model. For the purpose of certainty, it is assumed that an external magnetic field is in the first octant of the chosen coordinate system. The direction of the field is supposed to be the same throughout the volume of a sample and thus can be specified by unit vector ${\bf n} = \{ n_x, n_y, n_z \}$ with all $n_{\alpha} > 0$ ($\alpha = x, y, z$). If the field linearly increases along this direction, then ${\bf H} = {\bf n} ({\bf n} {\bf r}) h$ in the point with radius-vector ${\bf r}$. Here $h = \partial H_n / \partial r_n$ is the gradient of the field. The $k$-th rare-earth ion in the cell is under the action of the field,
    \begin{equation} \label{hkth}
        {\bf H}^{(k)} = {\bf n} ({\bf n} {\bf r}^{(k)}) h,
    \end{equation}
where ${\bf r}^{(k)}$ is the ion radius-vector, see Table \ref{tab1}.

It should be noted, that the rare-earth ions in iron garnets are also influenced by exchange field ${\bf H}_{\text{exch}} \sim \lambda {\bf M}_d$. Because of the magnetic anisotropy, the field is not collinear to the external ${\bf H}$ field and thus can contribute into the non-uniformity of the resulting magnetic field. However, the vector ${\bf M}_d$ deviates from the ${\bf H}$ vector in sufficiently strong external fields only slightly, thus it is possible to consider the actual inhomogeneous component of the magnetic field to be described by eq. \eqref{hkth} in the first approximation.

The components of the ${\bf H}^{(k)}$ field in eq. \eqref{hkth} can be expressed in the form
    \begin{equation} \label{hcom}
		H^{(k)}_{\alpha} = (h a/8) \cdot n_{\alpha} \cdot \beta^{(k)} ({\bf n}),
	\end{equation}
where $a$ is the edge length of the cell,
    \begin{equation} \label{bety} \begin{matrix}
	\beta^{(1)} ({\bf n})  = 6n_y + 3n_z, & \beta^{(2)} ({\bf n})  = 2n_y +  n_z, & \beta^{(3)} ({\bf n})  = 3n_x + 6n_z, \\
	\beta^{(4)} ({\bf n})  =  n_x + 2n_z, & \beta^{(5)} ({\bf n})  = 6n_x + 3n_y, & \beta^{(6)} ({\bf n})  = 2n_x +  n_y, \\
	\beta^{(7)} ({\bf n})  = 2n_y + 5n_z, & \beta^{(8)} ({\bf n})  = 6n_y + 7n_z, & \beta^{(9)} ({\bf n})  = 5n_x + 2n_z, \\
	\beta^{(10)} ({\bf n}) = 7n_x + 6n_z, & \beta^{(11)} ({\bf n}) = 2n_x + 5n_y, & \beta^{(12)} ({\bf n}) = 6n_x + 7n_y.
	\end{matrix} \end{equation}
From eq.~\eqref{eq08}, we find the expression for the dipole moment of the $k$-th ion,
    \begin{equation} \label{pulk} 
		{\bf P}^{(k)} = - M^{(k)}_{z} \left( C_1 H^{(k)}_{y} {\bf e}^{(k)}_{x} + C_2 H^{(k)}_{x} {\bf e}^{(k)}_{y} \right).
	\end{equation}
For the ferromagnetic phase (all the $M^{(k)}_{z}$ components are equal to $M$), the resulting dipole moment
    \begin{equation} \label{psum}
		{\bf P} = \sum_k {\bf P}^{(k)}.
	\end{equation}
From eqs. \eqref{pulk} and \eqref{psum} we finally obtain for dysprosium aluminium and holmium-yttrium iron garnets,
    \begin{equation} \begin{matrix}
		P_x = -\frac{1}{4} M a h (C_1 + C_2) \left( 2 n_x^2 - n_x n_y - n_x n_z \right), \\
		P_y = -\frac{1}{4} M a h (C_1 + C_2) \left( 2 n_y^2 - n_y n_x - n_y n_z \right), \\
		P_z = -\frac{1}{4} M a h (C_1 + C_2) \left( 2 n_z^2 - n_z n_x - n_z n_y \right].
	\end{matrix} \end{equation}
The polarization is estimated to be $10^2$ $\mu$C/m$^2$ if the magnetic field gradient reaches values of $10^3$~Oe/$\mu$m, which are actual for magnetic nanostructures.

To conclude, the inhomogeneous magnetic field results in the garnet crystal polarization depending on the field direction. It is interesting, that if ${\bf n} || [111]$, then $P_x = P_y = P_z = 0$, because $n_x = n_y = n_z = \frac{1}{\sqrt{3}}$.

It should also be noted, that the influence of an inhomogeneous magnetic field on magnetoelectric properties is very important for nanostructures and domain boundaries, for example, of the Neel type. In such systems, a field of $10^3$ -- $10^4$ Oe can be realized on a 10 nm scale, thus yielding enormous gradients up to $10^{10}$ Oe/cm and $P \sim 10^4$ -- $10^5$ $\mu$C/m$^2$.

Magnetoelectricity of rare-earth garnets can also be influenced by the flexomagnetism, see Refs.~\cite{ZP09} and \cite{Glin}, but the realization of the effects is owed to the inhomogeneity of the effective magnetic field a rare-earth ion is subjected to. The inhomogeneity is caused by the existence of domain boundaries in the subsystem of the iron ion magnetic moments.


\section{Linear magnetoelectric effects in epitaxial magnetic garnet films} \label{sec7}

The zero polarization of bulk rare-earth iron garnet single-crystals is owed to the equiprobable occupancy of the dodecahedral $c$-positions by rare-earth ions, ${\bf P} = \sum_{k = 1}^{12} {\bf P}^{(k)} = 0.$ In the case of garnet films, the situation changes dramatically. For example, in the growth process of (111) films there occurs a difference between the occupancies of the 1--6 and 7--12 positions (according to Eschenfelder \cite{Esch}, the X$_1$ and X$_2$ positions, relatively). This is the reason for uniaxial anisotropy in iron garnet films, which must result in the polarization occurrence in them.

In order to treat the phenomenon in a clear and simple way, we consider a so-called free-standing film approach and choose the ($\tilde{x}$,~$\tilde{y}$,~$\tilde{z}$) coordinate system with the $\tilde{z}$-axis directed along [111], ${\bf e}_{\tilde{x}} = [11\bar{2}]$, ${\bf e}_{\tilde{y}} = [1\bar{1}0]$, and ${\bf e}_{\tilde{z}} = [111]$. It is easy to show that the magnetoelectric energy of a primitive cell
    \begin{equation} \label{sume} 
		E_{me} = \sum_{k = 1}^{12} E^{(k)}_{me} = \varepsilon \mu_z (C_1 - C_2) \sqrt{3} \left[ {\bf H}_{\text{eff}} \times {\bf E} \right]_{\tilde{z}},
	\end{equation}
where the $E^{(k)}_{me}$ energies are given by eq.~\eqref{eq08}. Hence, for the (111) film, the polarization
    \begin{equation} \label{film} 
		{\bf P} = - \frac{\partial E_{me}}{\partial {\bf E}} = \varepsilon \mu_z (C_1 - C_2) \sqrt{3}
		\left( H_{\text{eff},\tilde{y}} \cdot {\bf e}_{\tilde{x}} - H_{\text{eff},\tilde{x}} \cdot {\bf e}_{\tilde{y}} \right),
	\end{equation}
where $\varepsilon$ is the difference between the occupation probabilities of the 1--6 and 7--12 positions. Note, that the ${\bf P}$ vector lies in the plane of the film and is perpendicular to the ${\bf H}_{\text{eff}}$ vector of the effective field.

It should also be noted, that there exists one more possibility of a polarization occurrence in iron garnet films. Because the structures of a film and a substrate are not completely coinciding and because there is inhomogeneity in crystal structure of a film (as a film grows, the rare-earth ions occupy the crystallographic positions that are slightly different from equilibrium positions), there occurs a mechanical tension gradient, which will be characterized by vector ${\bf K}$. This results in the new energy invariants such as $$ I \sim \sum_{k = 1}^{12} E^{(k)}_z K^{(k)}_x H^{(k)}_x M^{(k)}_z $$ and others like this, which can be obtained by replacements $x \rightleftarrows y$ and ${\bf E} \rightleftarrows {\bf K}$. From these invariants we infer for paramagnetic structures that polarization $P^{(k)}_i \sim \left( {\bf K}^{(k)} {\bf H}^{(k)} - K^{(k)}_i H^{(k)}_i \right) \eta^{9,1}_i$. Here $i = 1,2,3$ stands for $x, y, z$, relatively. The $\eta^{9,1}_i$ modes are given in Table II.

It should also be mentioned, that there exists another mechanism of the electric polarization occurrence in the films due to the mismatch between the structures of a film and substrate. The mechanism is considered in Ref.~\cite{Glin}.


\section{Conclusion}

During the last decade, the revival of a magnetoelectric effect has passed before our eyes the stage of renaissance and now turned into the very real boom. Garnets, as non-traditional materials towards magnetoelectricity, have been beyond the mainstream of the intensive research. However, our analysis shows that these materials can be promising candidates to observe a wide range of remarkable effects, namely, the traditional linear magnetoelectric effect (in Mn$_3$Al$_2$Si$_3$O$_{12}$), the piezoelectric effect induced by magnetic fields (in Nd$_3$Ga$_5$O$_{12}$), the piezomagnetic effect (in Dy$_3$Al$_5$O$_{12}$), etc. All the linear effects are magnetic field odd. It is also noteworthy, that external and internal effective $f$-$d$ exchange magnetic fields can form antiferroelectric structures, which coexist odd magnetic structures, for example in Dy$_3$Al$_5$O$_{12}$ and Ho$_x$Y$_{3-x}$Fe$_5$O$_{12}$, relatively.

Other fascinating aspects in the magnetoelectricity of garnets are the possibilities to realize the magnetoelectric effects in epitaxial thin films of those crystals in which the effects are forbidden by the symmetry and also in bulk garnets influenced by an inhomogeneous magnetic field. The latter aspect is especially important because it can launch new investigations on magnetoelectric properties of domain boundaries, which are underlain by inhomogeneities of internal magnetic fields caused by the boundaries themselves, without the implementation of the Dzyaloshinskii-Moriya mechanism.

Of course, all these issues require further detailed analysis and careful systematic study to overcome the limits of the symmetry considerations. It is impossible to cover all points in the magnetoelectricity of the vast family of garnets in a single journal paper. Our main aim is to draw attention to the magnetoelectricity of garnets, because the results obtained in this work give grounds to believe that there will be new advances and new discoveries on the way of further research. \bigskip

We wish to acknowledge the financial support of the Russian Ministry of Education and Science and 50 Labs Initiative of Moscow Institute of Physics and Technology.



\begin{thebibliography}{99}

\bibitem{Fieb} M. Fiebig, \textit{Revival of the Magnetoelectric Effect}, Journal of Physics D: Applied Physics, \textbf{38}, R123 (2005). 

\bibitem{Smol} G.A. Smolenskii and I.E. Chupis, \textit{Ferroelectromagnets}, Sov. Phys. Usp., \textbf{25}, 475 (1982). 

\bibitem{ZP12} A.P. Pyatakov and A.K. Zvezdin, \textit{Magnetoelectric and multiferroic media}, Phys. Usp. \textbf{55}, 557 (2012). 

\bibitem{Turo} E.A. Turov and V.V. Nikolaev, \textit{New physical phenomena caused by magnetoelectric and antiferroelectric interactions in magnets}, Phys. Usp., \textbf{48}, 431 (2005). 

\bibitem{Barj} V.G. Bar'yakhtar, V.A. L'vov, and D.A. Yablonskii D. A., \textit{Inhomogeneous magnetoelectric effect}, JETP Letters, \textbf{37}, 673 (1983). 

\bibitem{ZP09} A.K. Zvezdin and A.P. Pyatakov, \textit{Inhomogeneous magnetoelectric interaction in multiferroics and related new physical effects}, Phys. Usp., \textbf{52}, 845 (2009). 

\bibitem{Most} M. Mostovoy, \textit{Ferroelectricity in Spiral Magnets}, Phys. Rev. Lett. \textbf{96}, 067601 (2006). 

\bibitem{Logg} A.S. Logginov, G.A. Meshkov, A.V. Nikolaev, and A.P. Pyatakov, \textit{Magnetoelectric control of domain walls in a ferrite garnet film}, JETP Letters, \textbf{86}, 115 (2007). 

\bibitem{Yamg} T. Yamaguchi and K. Tsushima, \textit{Magnetic symmetry of rare-earth orthocromites and orthoferrites}, Phys. Rev. B \textbf{8}, 5187 (1973). 

\bibitem{Saha} R. Saha, A. Sundaresan, and C.N.R. Rao, \textit{Novel features of multiferroic and magnetoelectric ferrites and chromites exhibiting magnetically driven ferroelectricity}, Mater. Horiz. \textbf{1}, 20 (2014). 

\bibitem{Raje} B. Rajeswaran, D.I. Khomskii, A.K. Zvezdin, C.N.R. Rao, and A. Sundaresan, \textit{Field-induced polar order at the Neel temperature of chromium in rare-earth orthochromites: interplay of rare-earth and Cr magnetism}, Phys. Rev. B \textbf{86}, 214409 (2012). 

\bibitem{ZV08} A.K. Zvezdin and A.A. Mukhin, \textit{Magnetoelectric interactions and phase transitions in a new class of multiferroics with improper electric polarization}, JETP Letters, \textbf{88}, 505 (2008). 

\bibitem{DyFe} Y. Tokunaga, S. Iguchi, T. Arima, and Y. Tokura, \textit{Magnetic-field-induced ferroelectric state in} DyFeO$_3$, Phys. Rev. Lett. \textbf{101}, 097205 (2008). 

\bibitem{GdFe} Y. Tokunaga, N. Furukawa, H. Sakai, Y. Taguchi, T. Arima, and Y. Tokura, \textit{Composite domain walls in a multiferroic perovskite ferrite}, Nat. Mater. \textbf{8}, 558 (2009). 

\bibitem{SmFe} J.-H. Lee, Y.K. Jeong, J.H. Park, M.-A. Oak, H.M. Jang, J.Y. Son, and J.F. Scott, \textit{Spin-canting-induced improper ferroelectricity and spontaneous magnetization reversal in} SmFeO$_3$, Phys. Rev. Lett. \textbf{107}, 117201 (2011). 

\bibitem{Paol} \textit{Physics of magnetic garnets}, edited by A. Paoletti, North-Holland, Amsterdam (1978). 

\bibitem{Kric} B.B. Krichevtsov, V.V. Pavlov, and R.V. Pisarev, \textit{Giant linear magnetoelectric effect in garnet ferrite films}, JETP Letters, \textbf{49}, 535 (1989). 

\bibitem{Pisa} R.V. Pisarev,  B.B. Krichevtsov, V.N. Gridnev, V.P. Klin, D. Frohlich, and Ch. Pahlke-Lerch, \textit{Optical second-harmonic generation in magnetic garnet thin films}, J. Phys.: Condens. Matter, \textbf{5}, 8621 (1993). 

\bibitem{Pavl} V.V. Pavlov, R.V. Pisarev, A. Kirilyuk, and Th. Rasing, \textit{Observation of a transversal nonlinear magneto-optical effect in thin magnetic garnet films}, Phys. Rev. Lett. \textbf{78}, 2004 (1997). 

\bibitem{Yams} Y. Yamasaki, Y. Kohara, and Y. Tokura, \textit{Quantum magnetoelectric effect in iron garnet}, Phys. Rev. B \textbf{80}, 140412 (2009). 

\bibitem{Koha} Y. Kohara, Y. Yamasaki, Y. Onose, and Y. Tokura, \textit{Excess-electron induced polarization and magnetoelectric effect in yttrium iron garnet}, Phys. Rev. B \textbf{82}, 104419 (2010). 

\bibitem{Kaby} A.F. Kabychenkov, F.V. Lisovskii, and E.G. Mansvetova, \textit{Magnetoelectric effect in garnet films with the induced magnetic anisotropy in a nonuniform electric field}, JETP Letters, \textbf{97}, 265 (2013). 

\bibitem{Ogaw} H. Ogawa, E. Kita, Y. Mochida, K. Kohn, S. Kimura, A. Tasaki, and K. Shiratori, \textit{A low temperature phase transition in yttrium iron garnet}, J. Phys. Soc. Jpn. \textbf{56}, 452 (1987). 

\bibitem{VMCH} D. Varshalovich, A. Moskalev, and V. Khersonskii, \textit{Quantum theory of angular momentum} (World Scientific, Singapore, 1989). 

\bibitem{2013} A.I. Popov, D.I. Plokhov, and A.K. Zvezdin, \textit{Quantum theory of rare-earth multiferroics: Nd, Sm, and Eu ferroborates}, Phys. Rev. B \textbf{87}, 024413 (2013). 

\bibitem{Book} A.K. Zvezdin, V.M. Matveev, A.A. Mukhin, and A.I. Popov, \textit{Rare-earth ions in magnetically ordered crystals} (Nauka, Moscow, 1985), in Russian. 

\bibitem{Babu} G.A. Babushkin, V.A. Borodin, V.D. Doroshev, A.K. Zvezdin, R.Z. Livitin, and A.I. Popov, \textit{Magnetic phase transitions in samarium iron garnet and hypothesis of Ising ordering}, JETP Lett. \textbf{35}, 34 (1982). 

\bibitem{Nekv} V. Nekvasil, \textit{Magnetocrystalline anisotropy of samarium garnets}, Czech. J. Phys. \textbf{34}, 1052 (1984). 

\bibitem{ZV87} N.F. Vedernikov, A.K. Zvezdin, R.Z. Levitin, and A.I. Popov, \textit{Magnetic linear birefringence of rare-earth garnets}, Sov. Phys. JETP, \textbf{66}, 1233 (1987). 

\bibitem{Kova} V. Kovalev, \textit{Irreducible representations of the space groups}, Gordon and Breach, New York (1965). 

\bibitem{Izyu} Yu.A. Izyumov, V.I. Naish, and R.P. Ozerov, \textit{Neutron diffraction of magnetic materials}, Springer (1991). 

\bibitem{Wolf} W.P. Wolf, B. Schneider, D.P. Landau, and B.E. Keen, \textit{Magnetic and thermal properties of dysprosium aluminium garnet. Characteristic parameters of an Ising antiferromagnet}, Phys. Rev. B \textbf{5}, 4472 (1972). 

\bibitem{Gior} N. Giordano and W.P. Wolf, \textit{Induced staggered magnetic fields in antiferromagnets: microscopic mechanisms}, Phys. Rev. B \textbf{21}, 2008 (1980). 

\bibitem{Kolm} N.P. Kolmakova and A.I. Popov, \textit{Phase diagram of Dy$_3$Al$_5$O$_{12}$ and magnetic linear birefringence}, Physica B \textbf{162}, 71 (1990). 

\bibitem{Dill} J.F. Dillon, J.P. Remeika, and C.R. Staton, \textit{Linear magnetic birefringence in the ferrimagnetic garnets}, J. Appl. Phys. \textbf{41}, 4613 (1970). 

\bibitem{Nova} P. Novak, V. Nekvasil, T. Egami, P.J. Flanders, E.M. Gyorgy, L.G. van Uitert, and W.H. Grodkiewicz, \textit{Field induced magnetic moment and anisotropy of HoAG and TbAG}, J. Magn. Magn. Mater. \textbf{22}, 35 (1980). 

\bibitem{Levi} R.Z. Levitin, A.I. Popov, and V.V. Snegirev, \textit{Ferromagnetic resonance in holmium-yttrium ferrite-garnets in the Ising model}, Soviet Physics: Solid State, \textbf{24}, 3138 (1982). 

\bibitem{Plis} V.I. Plis and A.I. Popov, \textit{Linewidth of ferromagnetic resonance in holmium yttrium garnet ferrites}, Physics of the Solid State, \textbf{51}, 1176 (2009). 

\bibitem{Hast} J.M. Hastings, L.M. Corliss, and C.G. Windsor, \textit{Antiferromagnetic structure of dysprosium aluminum garnet}, Phys. Rev. \textbf{138}, 176 (1965). 

\bibitem{Glin} M.D. Glinchuk, A.V. Ragulya, and V.A. Stephanovich, \textit{Nanoferroics}, Springer Series in Material Science, Vol. 177, Springer Science + Business Media, Dordrecht (2013). 

\bibitem{Esch} A.H. Eschenfelder, \textit{Magnetic bubble technology}, Springer-Verlag Berlin Heidelberg New York (1981). 

\end{thebibliography}
\end{document}